\documentclass[a4paper]{article}
\usepackage{graphicx}
\usepackage{natbib}
\usepackage{graphics}
\newcommand{\tablenotemark}[1]{${}^{#1}$}
\newcommand{\tablecomments}[1]{}
\newcommand{\tablenotetext}[2]{${}^{#1}${#2}}

\begin{document}


\title{No Evidence of Time Dilation in Gamma-Ray Burst Data}
\author{David F. Crawford\\
Sydney Institute for Astronomy, School of Physics, University of Sydney\\
Address: 44 Market St, Naremburn, 2065, NSW, Australia.\\ davdcraw@bigpond.net.au}
\maketitle

\begin{abstract}
Gamma-Ray Bursts have been observed out to very high redshifts and provide time measures that are directly related to intrinsic time scales of the burst. Einstein's theory of relativity is quite definite that if the universe is expanding then the observed duration of these measures will increase with redshift. Thus gamma-ray burst measures should show a time dilation proportional to redshift. An analysis of gamma-ray burst data shows that the hypothesis of time dilation is rejected with a probability of 4.4$\times10^{-6}$ for redshifts out to z=6.6. Traditionally the lack of an apparent time dilation has been explained by an inverse correlation between luminosity and time measures together with strong luminosity selection as a function of redshift. It is shown that the inverse correlation between luminosity and some time measures is confirmed, but using concordance cosmology strong luminosity selection cannot be achieved. It may be possible to explain the apparent lack of time dilation with a combination of gamma-ray burst selection, some luminosity evolution and some time measure evolution. But this requires a remarkable coincidence in order to produce the apparent lack of time dilation. However the data are consistent with  a static cosmology in a non-expanding universe.

keywords(cosmology: observations, large-scale structure of universe, theory,  gamma rays: bursts)
\end{abstract}

\section{Introduction}
\label{s1}

Gamma-Ray bursts (GRB) are transient events with time scales of the order of seconds and with energies in the X-ray or gamma-ray region. \citet{Piran04} provides (a mainly theoretical) review and \citet*{Bloom03} give a review of observations and analysis. Although the reviews by \citet{Meszaros06} and \citet{Zhang07} cover more recent research and provide extensive references they are mainly concerned with GRB models. The only other objects for which time dilation has been observed are type 1a supernovae \citep{Goldhaber01, Foley05}. However this time dilation claim has been questioned by \citep{Crawford06, Crawford09a, Crawford09b}. Although the supernovae observations are clearly important in choosing between cosmologies they are not relevant to whether GRB show the effects of time dilation. This paper considers only the direct GRB observations and makes no assumptions about GRB models.

The search for the time dilation signature in data from GRB has a long history and before redshifts were available \citet{Norris94, Fenimore95a, Davis94} claimed evidence for the time dilation effect by comparing dim and bright bursts. However \citet{Mitrofanov96} found no evidence for time dilation. \citet{Lee00} found rather inconclusive results from a comparison between brightness measures and timescale measures. They also provide a brief summary of earlier results. Once redshifts became available \citet{Chang01} and \citet*{Chang02} using a Fourier energy spectrum method and \citet{Borgonovo04} using an autocorrelation method claim evidence of time dilation. The standard understanding, starting with \citet{Norris02} and \citet{Bloom03}, is that time dilation is present but because of an inverse relationship between luminosity and time measures it cannot be seen in the raw data. Because a strong luminosity-dependent selection produces an average luminosity that increases with redshift there will be a simultaneous selection for time measures that decrease with redshift which can cancel the effects of time dilation.

Here it is argued that there is no evidence for strong luminosity selection. Alternately the strong luminosity-redshift dependence may be due to luminosity evolution. In this case there is an increase in the average luminosity with redshift and not a selection of more luminous GRB. Consequently those time measures that show a strong relationship with luminosity must have evolved in a similar manner. Although it is possible that a combination of luminosity selection, selection of GRB by other characteristics and evolution may be sufficient to cancel time dilation it does require a fortuitous coincidence of these effects to completely cancel time dilation in the raw data. Another explanation is that the universe is not expanding and thus there is no time dilation. Not only is it shown that the data are consistent with a static cosmology but it is also shown that if a static cosmology is valid, it can readily explain the results from a concordance cosmology analysis.

The structure of this paper is to analyse recent GRB data to verify that there is no indication of time dilation in the raw time measures. Since the necessity of having an observed redshift makes this a specially selected sample of GRB the next step is to confirm that it shows the well known inverse relationship between luminosity and some time measures. It is also found, in agreement with the suggestion by \citep{Frail01, Bloom03}, that the data are consistent with the average energy of the GRB being constant. Since the dependence of the luminosity on redshift is strong  it is necessary to see whether it is due to selection or evolution. The next step is to investigate the selection process where it is shown that the data are inconsistent with strong luminosity-selection as a function of redshift. Then the alternative of  luminosity evolution is considered.  Next direct selection that depends on other characteristics of the GRB that could partially cancel the time dilation is considered. Although none of them is individually sufficient there is a possibility that a combination could be sufficient. Finally the data are shown to be in agreement with a static cosmology.

Recently \citet{Schaefer07a} has provided an excellent analysis of GRB relationships and provides tables of burst parameters of 69 bursts observed with redshift parameter z varying from 0.17 to 6.6. This redshift range is greater than that for type 1a supernovae for which there is also intrinsic timing information. Of relevance here are  $\tau_{lag}$ the lag time between a band of high energy gamma-rays and a band of lower energy gamma-rays, and $\tau_{RT}$ the shortest time over which the GRB light curve rises by half the peak flux of the pulse. A further measure is $V$, which is a measure of the variability of the light curve. Crudely, it is the number of spikes per second. In practice it is a normalised estimate of the fluctuations in the light curve relative to a smoothed version of the light curve. Finally \citet{Gehrels07} provides estimates of the time span $T_{90}$ that contains 90\% of the counts. These four time measures are determined from the original gamma-ray observations and are independent of any model for the burst mechanism. It can be argued that since the physical processes that produce these time measures are not well understood we cannot use them to test for time dilation. However time dilation must apply to all time measures. Here we start with the simplest assumption is that the characteristics of the GRB are the same at all redshifts. Then we consider whether evolution can explain the results.

In an expanding universe, the raw time measures should show an average duration that is directly proportional to ($1+z$), and the variability should show an inverse relationship. Time dilation provides a much stronger test of universal expansion than the Hubble redshift of frequencies. Of many explanations for the Hubble frequency redshift, the only one that includes time dilation is universal expansion.

This analysis uses the current standard cosmology, concordance cosmology, and following \citet{Schaefer07a} the mass density $\Omega_M$ is taken to be 0.27 and $\Omega_\Lambda$ which is proportional to the cosmological constant, has a value of 0.73. In this work the Hubble constant $H_0$ is assumed to have the value of 70 km s$^{-1}$ Mpc$^{-1}$. Because of the relatively small numbers the arguments used here are often statistical and the paper starts with a brief description of how the uncertainties are treated. Since the expected effect of time dilation is that it is a linear function of ($1+z$) it is appropriate to determine the exponent of ($1+z$) that has the best fit to the data and see if it is compatible with unity.

\section{Uncertainties}
\label{s2}
Since the observations have a very large spread, it is desirable to use logarithms so that the dynamic range is reduced, scale factors are irrelevant and the influence of outliers is reduced. In addition many of the variations in GRB such as external pressure and chemical composition are multiplicative. Thus from the central limit theorem the logarithm of a variable is likely to be closer to a normal distribution that the variable itself.

Although the measurements provided by \citet{Schaefer07a} have measurement uncertainties, it is very obvious, as \citet{Schaefer07a} has noted, that their scatter is much greater than the uncertainties would suggest and this scatter is mainly due the intrinsically different properties of the individual GRB. Therefore it is most likely that measurement uncertainties are a poor guide to the accuracy of each observation, and thus it is inappropriate to assume that the overall uncertainty is proportional to the measurement uncertainties. Instead a simple model is adopted where the additional uncertainty due to the intrinsic scatter is made proportional to the actual measure. Thus the variance for a measurement $y_i$ is set equal to $\sigma_i^2$+$(\eta y_i)^2$ where $\sigma_i$ is the measurement uncertainty for that measurement, $y_i$ is its value and $\eta$ is a constant for all the $y_i$ measures. The values of $\eta$ are estimated by requiring that the residual $\chi^2$ after doing a linear regression of $\log(y)$ against $\log(1+z)$ is equal to the number of degrees of freedom. This regression is done in order to remove any cosmologically important systematic effects on the estimate of $\eta$. Because this new term dominates the variance this procedure is almost the same as using unweighted values in regressions with $\log(y)$.

When both inputs to a linear regression have significant uncertainties the method ($fitexy$ subroutine) promoted by \citet{Press07} is used. This method retains symmetry between the variables in that the results are independent of which variable is chosen as the independent variable. For a linear regression where the $y_i$ variable has an uncertainty $\sigma_i$ and the $x_i$ variable has an uncertainty $\epsilon_i$ the method minimises $\chi^{2}$ where
\[
\chi^2=\sum\limits\frac{(y_i-a-b\:x_i)^2}{\sigma^2_i+ (b \epsilon_i)^2}
\]
If the uncertainties in $x$ are  small enough this method reduces to the standard weighted regression. Finally a small adjustment is made to the estimated errors in the regression results to make the residual $\chi^{2}$ value equal to the number of residual degrees of freedom. In effect the uncertainties in the measurements are used only to determine the relative weights. The final scaling of the uncertainty in any derived result is determined by the residual $\chi^{2}$. Since in this case this procedure increases the magnitude of the uncertainties the result is to decrease the significance of any dependency between the data and time dilation. It has been argued that the intrinsic scatter of the time variables is too large to show a significant dependence on time dilation. If this is true and time dilation is present then there is no change in the value of the expected exponent but the uncertainty in the exponent will be large.

\section{Procedure and results}
\label{s3}
Since there appears to be a clear distinction between short GRB, with $T_{90}$ less than 2 seconds \citep{Piran04, Borgonovo04}, and long GRB, the analysis is restricted to long GRB. Data for $\tau_{lag}$, $V$, and $\tau_{RT}$ are taken from \citet{Schaefer07a} (his Table 4) and $T_{90}$ data come from \citet{Gehrels07}. This second data set provided an extra 39 GRB that had only $T_{90}$ measurements. There were two measurements that were rejected as being outliers at the $5\sigma$ level. These outliers were  $\tau_{lag}$ from GRB030528, and $\tau_{RT}$ from GRB050824. In addition GRB020903 with a redshift of 0.25 was rejected because it was extremely weak and had no time measures. Since GRB60116 has a photometric redshift and \citet{Tanvir06} note that it is close to the Orion Nebula where the extinction is unreliable it has been omitted.

\begin{figure}[!htb]
\includegraphics[width=\columnwidth]{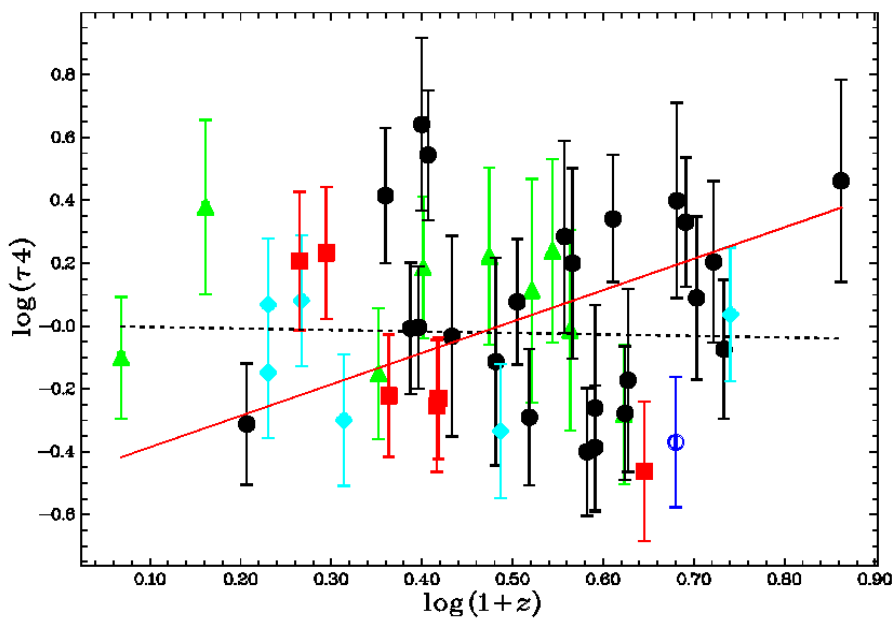}
\caption{Plot of the raw combined time measure  $\tau_{4}$  as a function of redshift. The dotted line is the line of best fit with a slope of -0.05$\pm$0.23. The solid line (in red) is the expected line for time dilation with a slope of one. The legend in Figure \ref{g2} shows the symbol and colour for each satellite detector which discovered that GRB. This legend is the same for all figures. \label{g1}}
\end{figure}

The estimation method is to use weighted linear regression, using the logarithms of the raw variables, to estimate the exponent as slope of the regression. Most of the regressions have $\log(1+z)$ as the independent variable. The estimates of the exponent for each of the four raw (i.e, uncorrected for time dilation) time  measurements are shown in Table \ref{grb1} together with the number of observations, the uncertainty (all uncertainties quoted are one-sigma values) in the exponent and the probability $p$ that the data are consistent with time dilation. The probability assumes a normal distribution for the exponents and is the probability that the observed exponent is greater than unity (or less than -1 for $V$).  In order to improve accuracy all four time measures were combined (as logarithms) into a new variable $\tau_{4}$ which is the weighted mean of the logarithms of $T_{90}$, $\tau_{lag}$, $1/V$, and $\tau_{RT}$. The weights (3.80, 2.63, 11.63, and 3.93 respectively) were the reciprocals of the average variances of their logarithms. To help avoid bias only bursts which had at least three measures were used to compute $\tau_{4}$. Although $\tau_{4}$ has no physical significance its use in this context is legitimate because the time dilation must apply to all the time measures. Figure \ref{g1} shows a plot of  $\tau_{4}$ as a function of redshift where the dashed line is the line of best fit.  The solid line shows the expected value with time dilation which is a power law with an exponent of unity. The $\chi^2$ value for an exponent of zero is 44.24 (46 DoF)
and for an exponent of unity it is 63.69 (46 DoF). \citet{Butler07} have done a different reduction of 218 Swift bursts of which 77 events have measured redshifts. Using only bursts with $T_{90}$ less than 2 seconds this data has an exponent for $T_{90}$ as a function of $(1+z)$ of $0.28\pm0.31$ in excellent agreement with the value in Table \ref{tbl-1}. The advantage of this data is that it is a homogeneous set derived from one satellite.

It is clear both from Figure \ref{g1} and Table \ref{grb1} that $\tau_{4}$ shows no dependence on redshift and that with an exponent of -0.04$\pm$0.23 there is a probability (one sided normal distribution) of 4.4$\times10^{-6}$ that the exponent is greater than or equal to unity. Although the central limit theorem predicts that the distribution of exponents should approximate the normal distribution it is in the tails of the distribution that we might expect some discrepancy. Thus the estimate 4.4$\times10^{-6}$, for the probability could be slightly incorrect which does not alter the conclusion that the probability that the raw time measures are consistent with time dilation is extremely unlikely.  In addition all the results are fully consistent with an exponent of zero. However it should be noted that $\tau_{lag}$ has a 9\% probability of being consistent with time dilation.

\begin{table}[!htb]
\begin{center}
\caption{Analysis of gamma-ray bursts for time dilation\label{tbl-1}}
\begin{tabular}{lccrc}
\hline
Variable &  $\eta$\tablenotemark{a} & N\tablenotemark{b} & Exponent\tablenotemark{c}  & Probability\tablenotemark{d} \\
\hline
$T_{90}$     & 0.51 & 84 &  0.28$\pm$0.28 & 4.7$\times10^{-3}$ \\
$\tau_{lag}$ & 0.55 & 36 &  0.18$\pm$0.61 & 8.9$\times10^{-2}$ \\
$V$          & 0.25 & 49 & -0.18$\pm$0.23 & 2.5$\times10^{-7}$ \\
$\tau_{RT}$  & 0.48 & 58 &  0.01$\pm$0.37 & 2.9$\times10^{-3}$ \\
$\tau_{4}$   & -    & 46 &  0.03$\pm$0.12 & 4.4$\times10^{-6}$ \\
\hline
\end{tabular}
\end{center}
\tablenotetext{a}{The value of $\eta$ for the logarithm of this variable.}\\
\tablenotetext{b}{The number of GRB used.}\\
\tablenotetext{c}{The exponent with respect to ($1+z$).}\\
\tablenotetext{d}{The probability of the exponent being greater than one (less than minus one for $V$).}\\
\end{table}

\section{Luminosity corrections}
\label{s4}
If there is time dilation a possible explanation \citep{Norris02, Bloom03, Schaefer07b} for its absence in the raw data is that the intrinsic time measures in a GRB are inversely correlated with the burst luminosity. The proposition is that because a strong luminosity-dependent selection produces an average luminosity that increases with redshift there will be a simultaneous selection for time measures that decrease with redshift. Clearly for this proposition to work the decrease in burst duration must closely cancel the increase due to time dilation. This proposition  can be tested by determining a correction (as an exponent) to the time measures that is a function of the luminosity. Then each GRB time measure is corrected by using its luminosity to get an estimate of the standard time measure that corresponds to a standard luminosity.

The precise method of correcting the time measures and determining the new exponents is as follows. Let $\tau_i$ be one of the time measures (where following standard procedure they are corrected for time dilation) for the $i$'th GRB and let $L_i$ be its bolometric luminosity. The correction exponent is determined by minimising the $\chi^2$ variate
\begin{displaymath}
\chi^2=\sum\limits\frac{(\log \left (\tau_i)- A - B\: \log(L_i) \right)^2} {\sigma^2_i+ (B\epsilon_i)^2},
\end{displaymath}
where $B$ is the required correction exponent, $\sigma_i$ is the uncertainty in $\log(\tau_1)$, $\epsilon_i$ is the uncertainty in $\log(L_i)$ and $A$ is the normalisation constant. Then $\tau_i L_i^{-B}$ is proportional to the time measure corrected to a standard luminosity. The next step is to determine the exponent as a function of redshift by minimising
\begin{displaymath}
\chi^2=\sum\limits\frac{\left (\log(\tau_i L_i^{-B})- a - b\: \log(1+z_i)\right)^2}{\sigma^2_i+ (b\epsilon_i)^2},
\end{displaymath}
where $\sigma_i$ is as before but now $\epsilon_i$ is the uncertainty in $\log(1+z_i)$ (which is negligible in this case). The correction coefficient $B$ and the new redshift dependent exponent $b$, are shown in Table \ref{grb2}. It is interesting that if the time measures were not corrected for time dilation the $B$ values, in the same order as in Table \ref{grb2}, are: $0.79\pm0.22$. $-1.11\pm0.16$, $0.66\pm0.15$, $-0.93\pm0.13$. The agreement between these two sets of exponents shows that the significant dependencies are intrinsic to the GRB and are not artifacts of the cosmology.  If the luminosity correction cancels the time dilation the expected exponent in column four is zero.
\begin{table}[!htb]
\begin{center}
\caption{Exponents corrected for luminosity dependence\label{grb2}}
\begin{tabular}{lcrr}
\hline\hline
Variable & Number   & Correction\tablenotemark{a} & Exponent\tablenotemark{b} \\
\hline
$T_{90}$    & 44&$ 0.72\pm0.23$ & -1.00$\pm$0.43 \\
$\tau_{lag}$& 36&$-1.15\pm0.15$ &  0.01$\pm$0.43 \\
$V$         & 49&$ 0.76\pm0.16$ &  0.48$\pm$0.30 \\
$\tau_{RT}$ & 58&$-1.02\pm0.15$ & -0.78$\pm$0.33 \\
$\tau_{4}$  & &                 & -0.61$\pm$0.22 \\ \hline
\end{tabular}
\end{center}
\tablenotetext{a}{The value of the luminosity correction exponent $B$, for this variable. }\\
\tablenotetext{b}{The exponent $b$, with respect to ($1+z$). It should be zero to show cancelled time dilation.}\\
\end{table}
They differ slightly from the values given by \citet{Schaefer07a} and \citet{Schaefer07b} because of the different methodology and a different use of uncertainties. Note that the time measure $T_{90}$ has a luminosity-correction  exponent of the wrong sign and that  $V$ has the expected opposite sign. The approximate inverse relationship between luminosity and time measures shown by earlier published results \citep{Norris02, Schaefer07b} is shown in the exponents for $\tau_{lag}$ and $\tau_{RT}$ but not by $V$ and $T_{90}$. It could be argued that this is because the two time measures $\tau_{lag}$ and $\tau_{RT}$ are more intrinsic to the nature of the burst spikes than the other two measures which include quiescent periods between spikes.

However there is an obvious trend in the luminosity corrections. The two time measures $\tau_{lag}$, and $\tau_{RT}$ which are directly related to the nature of the  burst spikes have luminosity dependence such that the product of the time measure times the luminosity is essentially constant. Since this product is proportional to the spike energy it implies that the burst energy is constant.  This is also in agreement with results from \citet{Panaitescu01, Frail01, Bloom03} who showed that the total energy determined by the afterglows is essentially constant. Furthermore this conclusion is similar to the results for supernovae \citep{Crawford06, Crawford09b} where it was found that the important constant characteristic of supernovae was their total energy and not their peak luminosity. Although a constant GRB energy rather than a constant luminosity is supported by these results the only significant implication of this concept is that it supports an exact inverse relationship between luminosity and burst-duration.

The proposition to be examined is that an inverse correlation between the burst luminosity and burst time measures in concordance cosmology can cancel time dilation. The measure of success of this proposition is to assess whether the computed exponents (as a function of ($1+z$)) are zero for each time measure after applying both time dilation and luminosity corrections to the raw values.  The final column in Table \ref{grb2} shows that the exponents for each time variables $\tau_{lag}$ and $V$ are consistent with the expected value of zero and $\tau_{RT}$ is marginally consistent with zero. Since $T_{90}$ has an exponent of -0.92$\pm$0.40 it is clearly anomalous. This is because although $T_{90}$ has  a small dependence on luminosity the change in luminosity is not sufficient to cancel time dilation for this time measure. Neither \citet{Norris02}, \citet{Bloom03}, nor \citet{Schaefer07b} discuss the $T_{90}$ time measure so that estimates of its calibration exponent or final redshift dependence cannot be compared with earlier results.

If we accept the proposition that the inverse relationship between luminosity and burst time measures is feasible there is still the problem that this  requires either strong luminosity selection or luminosity evolution. That there is a strong dependence of luminosity ($L_{bolo}$) on redshift is shown by the exponent for luminosity (using concordance cosmology) as a function of redshift which is 1.47$\pm$0.38. One possibility is that this exponent appears to be due to strong selection so that only the top of the GRB luminosity distribution is being sampled at large redshifts. It should be noted that this apparent strong selection depends also on the distance modulus used to convert from peak bolometric flux to luminosity. An alternative explanation to be considered in Section \ref{s5a} is that there  is luminosity evolution. Since neither of these possibilities can explain the lack of time dilation in the $T_{90}$ time measure, selection and evolution of time measures is considered in Section \ref{s6}.

\section{Luminosity selection without evolution}
\label{s5}
\begin{figure}[!htb]
\includegraphics[width=\columnwidth]{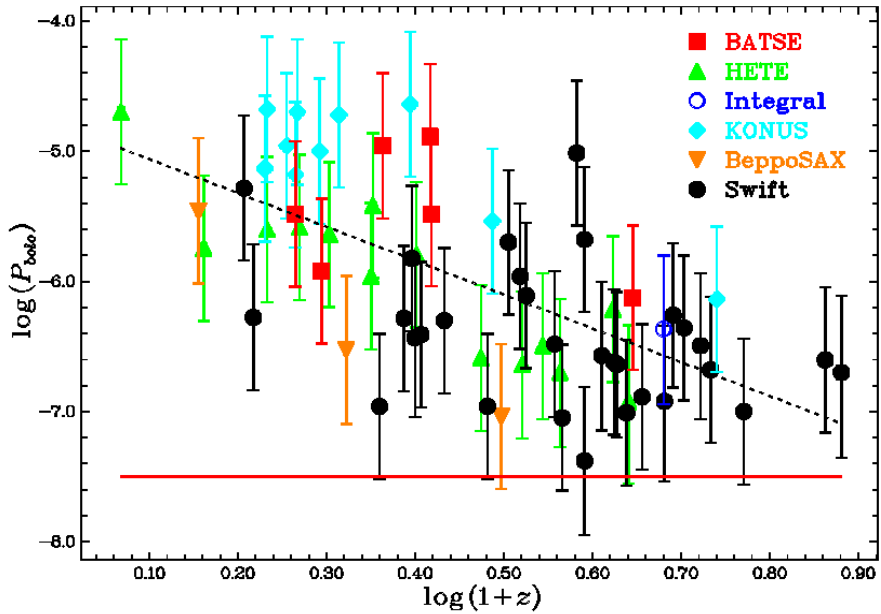}
\caption{Plot of the peak bolometric flux ($P_{bolo}$) in units of ergs cm$^{-2}$ s$^{-1}$ as a function of redshift. The dashed line is the line of best fit with a slope of -2.60$\pm$0.38. Although there is an apparent cutoff (shown in red as a solid line) at about $10^{-7.5}$ergs cm$^{-2}$ s$^{-1}$ there is no strong clustering just above the cutoff that would be expected if there was strong luminosity selection. The legend shows the symbol and colour for each satellite detector which discovered that GRB. \label{g2}}
\end{figure}

In this section it is shown that, in concordance cosmology, analysis of the selection process, assuming no evolution, shows that strong luminosity selection is unlikely.  Figure \ref{g2} shows a plot of the peak bolometric flux, which is derived from the observed gamma-ray spectrum (\citet{Schaefer07a}, equation 6), as a function of redshift. Although there is an apparent cut off in the peak bolometric flux at about $10^{-7.5}$ ergs cm$^{-2}$ s$^{-1}$ there is no evidence of strong clustering just above this cutoff. If there is strong luminosity selection and unless there is a lower limit to the luminosity distribution we would expect to see more GRB just above the cutoff at redshifts lower than about one.

If we knew the intrinsic density distribution of GRB as a function of luminosity one way to test the luminosity selection is to use a Monte Carlo method to distribute a random selection at constant volume density out to the highest redshift that was observed. Then to calculate the peak bolometric flux and reject all the members that are too faint to be observed. If the cosmology and the model are valid the remaining members should show (within statistical uncertainty) the same power law of luminosity as a function of redshift as the observations. There are two problems with this method. The first is that we do not have a good luminosity distribution and the second, which is more important is that we must allow for optical selection. All of these GRB have redshifts which were obtained from optical observations of the afterglow or of the host galaxy. Analysis of the number of GRB as a function of redshift (in either cosmology) shows that this optical selection is much more severe than the original GRB selection. One way to overcome these problems is to assume that the nearby GRB are a representative sample and then use their luminosities to compute the peak flux at higher redshifts and to investigate whether or not they would be detected. This was done with a simple Monte Carl procedure that starts with the set of 33 GRB that have redshift less than two. This gives a reasonable number of reference GRB without seriously biasing the sample to higher luminosity GRB. The problem of optical selection is overcome by using the redshifts of the observed GRB as a good measure of the redshift distribution. This method assumes that there is no significant correlation between gamma-ray and optical properties. In his review \citet{Piran04}  states that {\em there is no direct correlation between the $\gamma$-ray fluxes and the X-ray (or optical) afterglow fluxes}.

Then for each of the 65 GRB redshifts one of the 33 reference GRB was selected at random and from its luminosity a peak bolometric flux was computed for the new redshift. Next this selection was rejected if the new peak bolometric flux was less than $10^{-7.5}$ ergs cm$^{-2}$ s$^{-1}$. It might be argued that this bolometric flux cut-off is too simple and that the true selection criteria is a more complex function of time and gamma-ray energies. However here we are only interested in the major trend and minor variations in the selection criteria are not important. Another criticism is that the detection of a GRB is more accurately a function of the peak energy and not the peak bolometric flux. But the standard model is an inverse relationship  between the time measure and luminosity and not time measure and peak flux. Thus it is essential to show that there is strong luminosity selection for the standard model to succeed.

The random selection procedure was continued until there were 65 new luminosities for each of the 65 redshifts.  Table \ref{grb3} shows the average luminosity exponent for $10^3$ iterations. This simulation  was done for concordance cosmology (including time dilation corrections) and for a static cosmology (without time dilation corrections) (Section \ref{s7}). Clearly the simulated exponent for luminosity selection of $0.66$ for concordance cosmology is incompatible with the observed exponent of $1.47\pm0.38$.  However the exponents for the static cosmology are in good agreement.  Thus it is rather unlikely that the observed concordance cosmology luminosity exponent is due to luminosity selection.

\begin{table}[!htb]
\begin{center}
\centering \caption{Average luminosity exponent for concordance cosmology and for a static cosmology\label{grb3}}

\begin{tabular}{lccc}
\hline
& Number  & Concordance\tablenotemark{a}  & Static\tablenotemark{b} \\
\hline
Data                         &  65 & 1.47$\pm$0.38 & 0.38$\pm$0.21 \\
Simulation \tablenotemark{c} &  65 & 0.66$\pm$0.01 & 0.48$\pm$0.01 \\
Detector \tablenotemark{d}   &  65 & 0.35$\pm$0.01 & 0.30$\pm$0.01 \\
\hline
\end{tabular}
\end{center}
\tablenotetext{a}{The exponent for the observations using concordance cosmology.}\\
\tablenotetext{b}{The exponent for the observations using a static cosmology.}\\
\tablenotetext{c}{The exponent for the simulated data with luminosity selection.}\\
\tablenotetext{d}{The exponent for the simulated data with detector selection.}\\
\end{table}

\section{Peak energy selection}
\label{s5b}
\begin{figure}[!htb]
\includegraphics[width=\columnwidth]{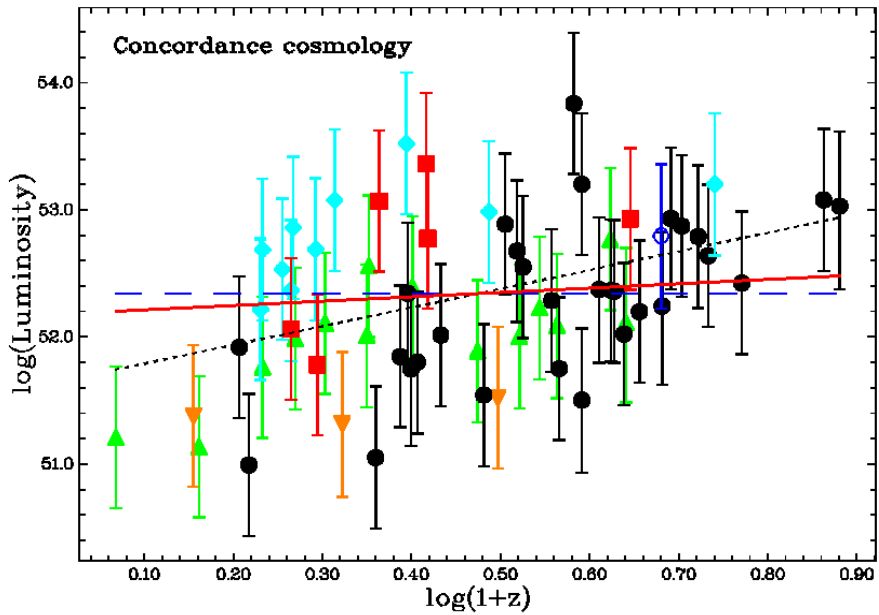}
\caption{Plot of the luminosity of GRB as a function of redshift for concordance cosmology. The dotted lines are the best fit power law to the densities (exponent 1.47$\pm$0.16).  The other lines show the power laws derived from the simulation analysis. The dashed (blue) line shows the power law for the luminosity selection simulation. The solid line (red) shows the power law for the detection efficiency. \label{g3}}
\end{figure}

Another selection process that may indirectly produce the large luminosity-redshift exponent (1.47$\pm$0.38) is a selection that is dependent on the peak photon energy ($E_{peak}$). For the Swift satellite the burst detection \citep{Markwardt07} requires a excess detector count rate (in the Burst Alert Telescope) over the background. Typically the threshold is set at 8-sigma. However the background is determined by a complex algorithm so that the detection depends on the peak bolometric flux  and it also depends on spike duration and spike rate. The other GRB telescopes \citep{Band93, Gorosabel01, Jager97, Kawai99} have a similar detection process. The detectors are essentially photon detectors so that their detection efficiency for a photon is almost independent of the photon energy provided it is within the energy window of the detector. If we consider the detection efficiency for a GRB placed at increasing redshifts the detection efficiency will be essentially constant until most of the photon energies fall below the lower limit of the detector's energy window. A rough estimate of the detection efficiency as a function of redshift is to assume that it is unity until the peak of the GRB photon energy spectrum reaches the lower energy cutoff of the detector. At higher redshifts the detection efficiency is zero. \citet{Schaefer07a} (his Table 2) provides both the upper and lower energy limits $E_{min}$ and $E_{max}$, for each GRB and the energy $E_{peak}$, at the peak of its spectrum. Using the same method as used for the simulation already described, the GRB with redshifts less than two were assumed to be a representative sample and the maximum redshift at which they could be observed was calculated. Then the number of surviving GRB were counted as a function of redshift to obtain a rough estimate of the selection efficiency. The exponent for a simulation of the detector selection is given in the last two rows of Table \ref{grb3}. The probability that the exponent of 0.35 is consistent with the data exponent of 1.47$\pm$0.38 is 1.5$\times 10^{-3}$ which shows that it is very unlikely that there is the required strong selection. Figure \ref{g3} shows the luminosity as a function of redshift for each GRB and their power law fit (dotted line). It also shows the simulated power law (dashed line), and the power law derived from the detection efficiency (solid line).

The distribution of luminosities in Figure \ref{g2} and two simulation results show that luminosity selection for the GRB is only a weak function of redshift and is clearly insufficient to produce the increase of average luminosity with redshift that is needed by the standard model.

\section{Luminosity evolution}
\label{s5a}
An alternative possibility to luminosity selection is that of luminosity evolution. After all in an expanding universe evolution is mandatory. The only objects that do not show evolution must have evolved to their current state early enough in the expansion to put their progenitors at redshifts beyond our current observations. Most of the work \citep{Piran04} on the evolution of GRB has been concerned with the nature and evolution of their host galaxy or on their volume density. Note that the approximately inverse relationship between luminosity and the time measures $\tau_{lag}$, $V$ and $\tau_{RT}$ implies that a strong luminosity evolution requires that all of these time measure must have strong evolution.

\citet*{Lloyd-Ronning02} using concordance cosmology find that there is luminosity evolution such that the luminosity is proportional to $L\propto (1+z)^{1.4\pm \approx0.5}$. More recently \citet{Kocevski06} find an exponent of $1.7\pm0.3$.  A direct fit of luminosity as a function of redshift for the current data has the exponent 1.47$\pm$0.16 (Table \ref{grb3}). Since \citet{Lloyd-Ronning02} used a method that allows for possible luminosity selection effects and the current analysis has shown that luminosity selection effects are small there is good agreement between the exponents. Thus the results in Table \ref{grb2} are  applicable to both luminosity selection and luminosity evolution with similar conclusions. That is this luminosity evolution can explain why time dilation is cancelled for $\tau_{RT}$ and $V$ and possibly for $\tau_{lag}$ but not for $T_{90}$.

However there is a major defect in using evolution to explain the cancellation of time dilation. The computation of luminosity requires a distance measure that is determined by the assumed cosmology (\citet{Schaefer07a}, equation 6). Thus it is possible that the apparent luminosity evolution could be due to an incorrect luminosity calculation. If in this case this 'evolution' is then used to provide the required luminosity-redshift dependence needed to cancel time dilation then it is invalid to use this result to justify the cosmology and the occurrence of time dilation. For example if the universe was static the major difference in the calculation of the luminosity is the removal of a factor of ($1+z$) that was there to allow for time dilation. Consequently the luminosity-redshift exponent would change from 1.47 to about 0.4. This would give a luminosity evolution of about $L\propto (1+z)^{0.4}$ and a small change in the time measures. Since in this case there is no time dilation everything is self consistent.

\section{Time measure selection}
\label{s6}
A further  possibility that could explain this disagreement is that there is redshift-dependent selection that depends on other characteristics of the GRB. The obvious characteristics are the time measures. Clearly a redshift selection based on the time measures could easily cancel time dilation and at the same time  produce the observed luminosity distribution.  However, such a selection, if strong enough, would produce a reduced spread of $\tau_{4}$ at large redshifts. Figure \ref{g1} shows  no such effect.

An important selection mechanism is the pulse width selection effect that comes from the characteristic of GRB in that they have narrow pulse widths at high gamma-ray energies but extended trains of broad pulses at low gamma-ray energies \citep{Piran04}. Thus at high redshifts the decrease of low gamma-ray energies to below the detector threshold will produce an increase in the observed intrinsic gamma-ray energies and hence a decrease in the pulse widths.  Because of the luminosity time measure dependence this selection could, in principle, produce the strong dependence of luminosity on redshift. \citet{Fenimore95b} and \citet{Dado07} show that the width of a GRB pulse scales with energy like $t_{FWHM}\propto E^{-0.43\pm0.10}$ where $E$ is the gamma-ray energy. \citet{Schaefer07a} provides values for the peak ($E_{peak}$) of  the gamma-ray distribution for each GRB. With their redshift correction these $E_{peak}$ values have an exponent of $0.80\pm0.25$ with respect to ($1+z$). Thus the pulse width at the peak of the energy distribution will have an exponent $-0.35\pm0.13$ as a function of redshift.  At the low energy extreme if  we assume that $E$ is the threshold energy of the detector then this translates to a pulse width that has an exponent of -0.41 with respect to ($1+z$). Since the average exponent will lie between these two values it is very unlikely that the redshift exponent of $\tau_{RT}$ (assuming that it is proportional to the pulse width) could have the value of minus one needed to cancel time dilation.

The effects of this pulse width selection on $\tau_{lag}$ are harder to evaluate. It is defined as the delay in the times of peaks between a hard band (100-300 keV) and a soft band (25-50 keV) after considerable data massaging in order to remove noise and artifacts \citep{Schaefer07a}. One effect that is predictable is the failure to detect the soft band and therefore to obtain a value for $\tau_{lag}$. For this sample the highest redshift which has a value for $\tau_{lag}$ is the Swift GRB060223 with $z$=4.41 which suggests that this failure is not very important for this data.  Since $\tau_{lag}$ appears to be an intrinsic property of the burst mechanism it is unlikely that its values are significantly modified by this pulse width selection process.

Although the omission of some lower energy pulses may decrease the variability measure other changes such as the decrease in pulse widths may increase it. Thus it is unlikely that $V$ treated as a time measure would show a strong redshift dependence. Since the $T_{90}$ measure is  the time it takes to accumulate from 5\% to 95\% of the total fluence of a burst it is reasonable that its value at high redshifts could be decreased by the omission of low energy pulses at the beginning and end of the burst but this reduction is unlikely to produce  sufficient modification to values of $T_{90}$. Since the soft band gamma-rays are detectable out to about  $z=4$,  $T_{90}$ must be reduced by a factor of $\approx 5$ for the time dilation to be cancelled. Although it is difficult to imagine some selection effect that could be as strong as this, it is possible that it is the result of some evolutionary effect.

Thus it is very likely that all of the time measures will show the effects of some selection but the only time measure that is expected to show a significant selection effect is $\tau_{RT}$ and as shown above it is very unlikely that it could have an exponent of minus one needed to cancel time dilation. The conclusion is that although time measure selection does occur it is not sufficient to cancel time dilation.

There remains the possibility that a combination of luminosity selection, pulse width selection and evolution could be sufficient to cancel time dilation. However, such a combination is strongly constrained by requiring that its effects almost exactly cancel time dilation in the raw time measures. That is each time measure must be approximately proportional to $(1+z)^{-1}$.

\section{Static cosmology}
\label{s7}
\begin{figure}[!htb]
\includegraphics[width=\columnwidth]{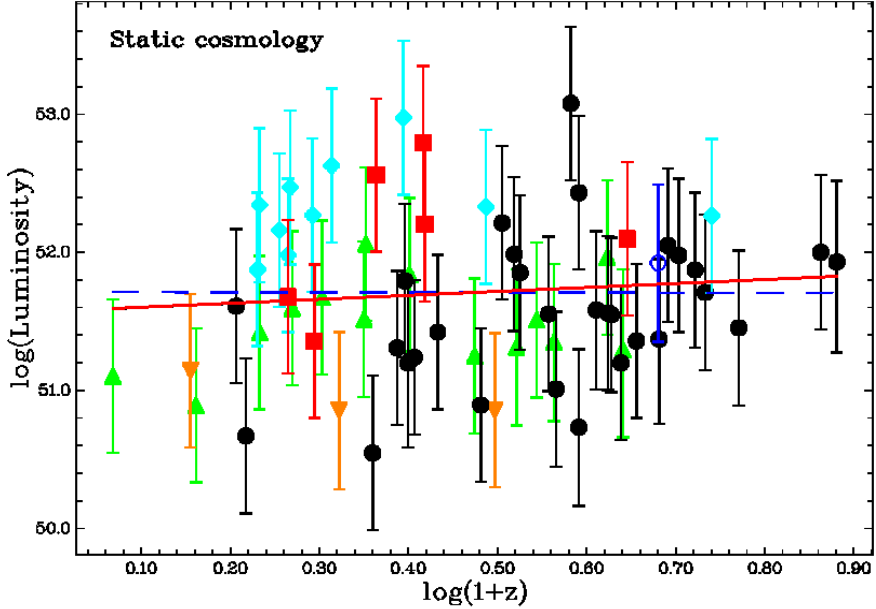}
\caption{Plot of the luminosity of GRB as a function of redshift for a static cosmology. The dotted lines are the best fit power law to the densities.  The dashed lines show the power laws derived from the simulation analysis. The dashed (blue) line shows the power law for the luminosity selection simulation. The solid line (red) shows the power law for the detection efficiency simulation.\label{g4}}
\end{figure}

Since concordance cosmology is based on the principle that the universe is expanding it cannot be directly used to test for universal expansion. Equally a static cosmology that is based on a non-expanding universe has the same problem. However a distinction can be made if the observations clearly favour one cosmology over the other. In a static universe the redshift must be due to some mechanism other than expansion. Here we assume, without  explanation, that the normal Hubble redshift-distance relationship holds. The main modification needed to concordance cosmology to simulate a static cosmology is to remove a ($1+z$) factor from the distance modulus. This factor arises because the flux is an energy rate and the rate is subject to time dilation. Rather than such an ad hoc modification, a static cosmology that agrees with other cosmological observations \citep{Crawford06, Crawford09a} is used instead. Apart from a scale factor the differences between its distance measures and that for a modified concordance cosmology are much smaller than the uncertainties in the data used here. The results using this static cosmology and exactly the same data and analysis (including the intrinsic luminosity time measure relationship) that was used for concordance cosmology are shown in Table \ref{grb3} and Figure \ref{g4}. Since the exponents shown in the right hand column of Table \ref{grb3} are consistent with zero this shows that the data are completely consistent with this static cosmology. As expected the average luminosity is essentially independent of redshift and whether not luminosity corrections are applied  makes very little difference. Naturally if there is no expansion all the results in Table \ref{grb1} are correct and do not require any explanation.

If a static cosmology is correct then the results for concordance cosmology are easily understood. For simplicity we assume that there is no luminosity selection then the inclusion of a ($1+z$) factor in the distance modulus will produce an apparent exponent of plus one in the luminosity distribution. Next applying the luminosity time measure relationship will make the apparent time measures have an exponent of minus one. Finally the inclusion of the time dilation will result in the time measures having zero exponent which is what is seen in Table \ref{grb1}.

\section{Conclusion}
The first conclusion is that the raw time measures showed no evidence of time dilation with the combined measure $\tau_4$ having a probability of 4.4$\times10^{-6}$ of occurring if time dilation is present. With the possible exception of $\tau_{lag}$ none of the time measures show a significant effect of time dilation. The standard explanation is that there is a luminosity time measure relationship such that selection by luminosity produces an effective selection of time measures that cancels the observation of time dilation. This relationship was confirmed for $V$ and $\tau_{RT}$ and since the luminosity has an exponent of 1.47$\pm$0.16 with respect to ($1+z$) and as shown in Table \ref{grb2} this is sufficient to cancel time dilation for these time measures but not for $T_{90}$. But it was shown that luminosity selection could not produce this strong dependence of luminosity on redshift but that luminosity evolution could. Examination direct selection of the time measures showed that except for $\tau_{RT}$ none of them would have strong enough selection and even  $\tau_{RT}$ would only have an exponent of $\approx -0.4$.  It is possible that a combination of luminosity evolution, time measure evolution  with some luminosity selection and some time measure selection could produce the required cancellation of time dilation. However this requires the coincidence that (except possibly for $\tau_{lag}$) all the time measures end up with an exponent with respect to ($1+z$) that is close to minus one.

Alternatively it is shown that all the data are consistent with a static cosmology and that if a static cosmology is valid then it can easily explain the results obtained from a concordance cosmology analysis. There is no strong dependence of luminosity on redshift in this static cosmology, it is an artifact of concordance  cosmology because it includes a factor of ($1+z$) to allow for time dilation.  Thus there is strong support for the notion that there is no time dilation that the universe is not expanding.

\section{Acknowledgments}
This research has made use of the NASA/IPAC Extragalactic Database (NED) that is operated by the Jet Propulsion Laboratory, California Institute of Technology, under contract with the National Aeronautics and Space Administration. The graphics have been done using the DISLIN plotting library provided by the Max-Plank-Institute in Lindau. I thank the referee for constructive criticisms.


\begin{thebibliography}{}
\bibitem[\protect\citeauthoryear{Band {\it et al.}}{1993}]{Band93} Band, D., Matteson, J., Ford, L., Schaefer, B., Palmer, D., Teegarden, B., Cline, T., Briggs, M., Paciesas, W., Pendleton, G., Fishman, G., Kouveliotou, C., Meegan, C., Wilson, R., 1993, Apj, 413, 281
\bibitem[\protect\citeauthoryear{Bloom, Frail \& Kulkarni}{Bloom {\it et al.}}{2003}]{Bloom03} Bloom, J. S., Frail D.A., Kulkarni S.R., 2003, ApJ, 594, 674
\bibitem[\protect\citeauthoryear{Borgonovo}{2004}]{Borgonovo04} Borgonovo, L., 1994, A\&A, 418, 487
\bibitem[\protect\citeauthoryear{Butler {\it et al.}}{2007}]{Butler07}Butler, N. R., Kocevski, D.,Bloom, J. S. \& Curtis, J. L., 2007, ApJ, 671, 656
\bibitem[\protect\citeauthoryear{Chang}{2001}]{Chang01} Chang, H. -Y, 2001, ApJ, 557, L85
\bibitem[\protect\citeauthoryear{Chang, Yoon \& Choi}{Chang {\it et al.}}{2002}]{Chang02} Chang, H. -Y., Yoon, S, -J., Choi, C. -S., 2002, A\&A, 383, L1
\bibitem[\protect\citeauthoryear{Crawford}{2006}]{Crawford06} Crawford, D. F., 2006, Curvature Cosmology, Boca Ratan, BrownWalker Press
\bibitem[Crawford(2009a)]{Crawford09a} Crawford, D. F., 2009a, (Contains second edition of the book 'Curvature Cosmology'), http://www.davidcrawford.bigpondhosting.com
\bibitem[\protect\citeauthoryear{Crawford}{2009b}]{Crawford09b} Crawford, D. F., 2009b, arXiv:0901.4172
\bibitem[\protect\citeauthoryear{Dado, Dar \& De R\'{u}jula}{Dado {\it et al.}}{2007}]{Dado07}Dado, A., Dar, A., De R\'{u}jula, A., 2007, ApJ, 663, 400
\bibitem[\protect\citeauthoryear{Davis {\it et al.}}{1994}]{Davis94} Davis, S. P., Norris, J. P., Kouveliotou, C., Fishman, C., Meegan, C. A., Pasiesas, W. C., 1994, in AIP Conf. Proc. 307, Gamma-Ray Bursts,ed. G. J. Fishman, J. J. Brainard \& K. C. Hurley, New York, AIP,182
\bibitem[\protect\citeauthoryear{Fenimore \& Bloom}{1995a}]{Fenimore95a} Fenimore, E. E., Bloom, J. S., 1995, ApJ, 453, 25
\bibitem[\protect\citeauthoryear{Fenimore {\it et al.}}{1995b}]{Fenimore95b}Fenimore, E. E., in't Zand, J. J. M., Norris, J. P., Bonnell, J. T. \& Nemiroff, R. J., 1995, ApJ, 448, L101
\bibitem[\protect\citeauthoryear{Foley {\it et al.}}{2005}]{Foley05}Foley, R. F., Filippenko, A. V., Leonard, D. C., Riess, A. G., Nugent, P. \& Perlmutter, S., 2005, ApJ, 626, L11
\bibitem[\protect\citeauthoryear{Frail {\it et al.}}{2001}]{Frail01} Frail, D. A., Kulkarni, S. R., Sari, R., Djorgovski, S. G., Bloom, J. S., Galama, T. J., Reichart, D. E., Berger, E., Harrison, F. A., Price, P. A., Yost, S. A., Diercks, A., Goodrich, R. W., Chaffee, F., 2001, ApJ, 562, L55
\bibitem[\protect\citeauthoryear{Gehrels}{2007}]{Gehrels07} Gehrels. N., 2007, http://heasarc.nasa.gov/\-docs/\-swift/\-results/
\bibitem[\protect\citeauthoryear{Goldhaber {\it et al.}}{2001}]{Goldhaber01}Goldhaber, G., Groom, D. E., Kim, A., Algering, G., Astier, P., Couley, A., Deustua, S. E., Ellis, R., Fabbro, S., Fruchter, A. S., Goobar, A., Hook, I., Orwin, M., Kim, M., Knop, R. A., Lidman, C., McMahon, R., Nugent, P. E., Pain, R., Panagia, N., Pennypacker, C. R., Perlmutter, S., Ruiz-Lapuente, P., Schaefer, B., Walton, N. A., York, T., 2001, ApJ. 558, 359
\bibitem[\protect\citeauthoryear{Gorosabelet {\it et al.}}{2001}]{Gorosabel01} Gorosabel, J., Lund, N., Brandt, S., Westergaad, N.J., 2002, in 4th INTEGRAL Workshop, Exploring the Gamma-Ray Universe, (ESA SP-459, Sep, 2001)
\bibitem[\protect\citeauthoryear{Jager {\it et al.}}{1997}]{Jager97} Jager, R., Mels, W. A., Brinkman, A. C., Galama, M. Y., Goulooze, H., Heise, J., Lowes, P., Muller, J. M., Naber, A., Rook, A., Schuurhof, R., Schuurmans, J. J., Wiersma, G., 1997,  Astron. Astrophys. Suppl. Ser., 236, 557
\bibitem[\protect\citeauthoryear{Jawai {\it et al.}}{1999}]{Kawai99} Kawai, N., Matsuoka, M., Yoshida, A., Shirasaki, Y., Namiki, M., Takagishi, K., Yamauchi, M., Hatsukade, I., Fenimore, E. E.,  Galassi, M., 1999, A\&AS. 138, 563
\bibitem[\protect\citeauthoryear{Kocevski \& Liang}{2006}]{Kocevski06}Kocevski, D. \& Liang, E., 2006, ApJ, 642, 371
\bibitem[\protect\citeauthoryear{Lee {\it et al.}}{2000}]{Lee00} Lee, A., Bloom, E. D. \& Petrosian, V., 2000, ApJ, 131,21
\bibitem[\protect\citeauthoryear{Lloyd-Ronning,  Fryer, \& Ramirez-Ruiz}{Lloyd-Ronning {\it et al.}}{2002}]{Lloyd-Ronning02} Lloyd-Ronning, N. M., Fryer, C. L., Ramirez-Ruiz, E., 2002, ApJ, 574, 554
\bibitem[\protect\citeauthoryear{Markwardt {\it et al.}}{2007}]{Markwardt07} Markwardt, C. B., Bathelmy, D. D., Cummings, J. C., Hullinger, D., Krimm, H. A., Parsons, A., 2007, The Swift BAT Software Guide,
    http://swift.gsfc.nasa.gov/\-docs/\-swift/\-analysis/\-bat\_swguide\_v6\_3.pdf
\bibitem[\protect\citeauthoryear{M\'{e}sz\'{a}ros}{2006}]{Meszaros06} M\'{e}sz\'{a}ros, P., 2006, Rep. Prog. Phys., 2259
\bibitem[\protect\citeauthoryear{Mitrofanov {\it et al.}}{1996}]{Mitrofanov96} Mitrofanov, I. G., Chernenko, A. M., Pozanenko, A. S., Briggs, M. S., Paciesas, W. S., Fishman, G. J., Meegan, C. A., Sagdeev, R. Z., 1996, ApJ, 459,570
\bibitem[\protect\citeauthoryear{Norris}{2002}]{Norris02} Norris, J. P., 2002, ApJ, 579, 386
\bibitem[\protect\citeauthoryear{Norris {\it et al.}}{1994}]{Norris94} Norris, J. P., Nemiroff, R. J., Scargle, J. D., Kouveliotou, C., Fishman, G. J., Meegan, C. A., Paciesas, W. S., Bonnell, J. T., 1994, ApJ, 424 540
\bibitem[\protect\citeauthoryear{Panaitescu \& Kumar}{2001}]{Panaitescu01} Panaitescu, A., Kumar, P., 2001, ApJ, 560, L49
\bibitem[\protect\citeauthoryear{Piran}{2004}]{Piran04} Piran, T., 2004, Rev. Mod. Phys, 76, 1143
\bibitem[\protect\citeauthoryear{Press {\it et al.}}{2007}]{Press07} Press, W. H., Teukolsky, A. A., Vetterling, W. T., Flannery, B. P., 2007, Numerical Recipes, 3rd edn., Cambridge, Cambridge University Press
\bibitem[\protect\citeauthoryear{Schaefer}{2007a}]{Schaefer07a} Schaefer, B. E., 2007 ApJ, 660, 16
\bibitem[\protect\citeauthoryear{Schaefer \& Collazzi}{2007b}]{Schaefer07b} Schaefer, B. E., Collazzi, A. C., 2007 ApJ, 656, L53
\bibitem[\protect\citeauthoryear{Tanvir {\it et al.}}{2006}]{Tanvir06}Tanvir, N. R., Levan, A. J., Pridday, R. S., Fruchter, A. S. \& Hjorth, J., 2006, GCN4602
\bibitem[\protect\citeauthoryear{Zhang}{2007}]{Zhang07} Zhang, B., 2007, Chin. J. Astron. Astrophys., 1

\end{thebibliography}
\end{document}